\documentclass[prl,amsfonts,amssymb,showpacs,floats,twocolumn,superscriptaddress,aps]{revtex4}

\usepackage{amsmath}
\usepackage{bm}
\usepackage{graphicx}

\def\punkt{\;\; .}

\def\w{\omega}
\def\H{{\cal H}}
\def\e{\epsilon}

\def\Tr#1{\textrm{Tr}\left[#1\right]}

\begin{document}

\title{On steady-state currents through nano-devices: a
  scattering-states numerical renormalization group approach to open
  quantum systems}

\author{Frithjof B. Anders}
\affiliation{Institut f\"ur Theoretische Physik, Universit\"at Bremen,
                  P.O. Box 330 440, D-28334 Bremen, Germany}

\date{June 13, 2008}

\begin{abstract}
We propose a  numerical renormalization group (NRG) approach to steady-state
currents through nano-devices. A discretization of the
scattering-states continuum ensures the correct boundary condition for
an open quantum system. 
We introduce two degenerate Wilson chains for current carrying left and
right-moving electrons reflecting time-reversal symmetry in the
absence of a finite bias $V$.
We employ the time-dependent NRG to evolve the
known steady-state density operator for a non-interacting
junction into the density operator 
of the fully interacting nano-device at finite bias. 
We calculate the differential conductance as function of $V,T$ and
the external magnetic field.
\end{abstract}

\maketitle

\paragraph*{Introduction:}
The  description of quantum systems out of equilibrium  is one the
fundamental challenges in theoretical physics. Even a simple
non-equilibrium situation,  the current transport through an interacting
junction at finite bias is not fully  understood.
The Coulomb blockade\cite{KastnerSET1992} and advent of the
experimental realizations of the Kondo effect in such
devices\cite{NatureGoldhaberGordon1998,Kouwenhoven2000} requires a
many-body description at low temperatures.

While the equilibrium dynamics is well
understood\cite{BullaCostiPruschke2007}, the  non-equilibrium  
steady-state has been mainly investigated using perturbative
approaches\cite{MeirWingreen1994,KoenigSchoeller98,RoschPaaskeKrohaWoelfe2003,GezziPruschkeMeden2007}
based on Keldysh theory\cite{Keldysh65},
the Toulouse point\cite{SchillerHershfield95} and the flow
equation\cite{Kehrein2005}. Landauer-Buettiker type
approaches\cite{MolecularElectronicsBook2005} treat the charging
effect only on a mean-field level by mapping the strongly interacting
quantum problem onto a model of non-interacting fictitious particles,
unsuitable to describe the Coulomb-blockade
physics\cite{KastnerSET1992}. In weak coupling and high temperature,
the ac and dc-transport through molecular wires can be addressed by a
quantum master equation for the reduced density matrix of the
junction\cite{welack-2006-124}. All those approaches have only a
limited validity of their parameter regimes. Recently, Han proposed an
alternative perturbative method\cite{JongHan2006} based on
Hershfield's steady-state density
operator\cite{Hershfield1993,HanHeary2007,Oguri2007,DoyonAndrei2005}.
Based on similar ideas, a scattering-states  Bethe-ansatz approach to
an  interacting spinless quantum dot has been
implemented\cite{MethaAndrei2005} for finite bias. 

We present a numerical renormalization group
approach\cite{BullaCostiPruschke2007} to open quantum systems based on
scattering states\cite{Hershfield1993}.
It combines 
(i)  Wilson chains for single-particle scattering states proposed below,
(ii) Hershfield's  steady-state density operator\cite{Hershfield1993}
for a non-interacting junctions at finite bias and
(iii) the  time-dependent NRG
(TD-NRG)\cite{AndersSchiller2005,AndersSchiller2006,AndersNeqGf2008}. 
Our scattering-states basis will be also useful for   Quantum
Monte Carlo and density matrix renormalization group (DMRG)
approaches\cite{SchollwoeckDMRG2005}. With our non-perturbative
method, steady-state currents through interacting nano-devices can  be
obtained accurately for arbitrary temperatures, magnetic fields and
interaction strength.

Dissipative steady-state currents only occur in {\em open quantum
  system} in which the system size $L$ has been sent to $L\to\infty$ before  $t\to\infty$.
Transient currents can be calculated on a finite-size
tight-binding chain within the TD-NRG as well as the
time-dependent DMRG\cite{Schmitteckert2004,SchollwoeckDMRG2005}. 
However, such transient currents  vanish  for $t\to\infty$ or
even reverse their  sign\cite{Schmitteckert2004} in those approaches, a
consequence of the non-interchangeable limit  $t\to\infty$ and
$L\to\infty$\cite{DoyonAndrei2005}. We circumvent this problem by
discretizing a  single-particle scattering states basis. Therefore,
those states remain current carrying and a  faithful representation of
an {\em open quantum system}.

\paragraph*{Theory:}

Interacting quantum dots (QD), molecular junctions or other nano-devices
are modelled by the interacting region $\H_{imp}$, a set of
non-interacting reservoirs $\H_{B}$ and a coupling between both
sub-system $\H_I$: $\H = \H_{imp} + \H_{B} + \H_{I}$.
Throughout this paper, we restrict ourselves to a junction with a single
spin-degenerate orbital $d$ with energy $E_d$, subject to an external
magnetic field $H$ and an on-site Coulomb repulsion $U$. The
orbital is coupled to a left (L) and a right (R) lead
via the  tunneling matrix elements $V_{\alpha=L,R}$, and $\H$ given by
\begin{eqnarray}
\label{eqn:siam}
{\cal H} &=& 
\sum_{\sigma \alpha=L,R} \int d\e \, \e \, c^\dagger_{\e\sigma\alpha}
c_{\e\sigma\alpha}
 \\
&& 
             + \sum_{\sigma = \pm 1}
                   \left[
                        E_d +\frac{U}{2} - \frac{\sigma}{2} H
                   \right ]
                   \hat{n}^d_\sigma
+ \frac{U}{2}\left( \sum_{\sigma} \hat{n}^d_\sigma -1\right)^2
\nonumber \\
 && + \sum_{\alpha\sigma}   V_{\alpha} \int d\e \,\sqrt{\rho(\e)}
               \left\{
                    d^\dagger_\sigma c_{\e\sigma\alpha} +
                    c^\dagger_{\e\sigma\alpha} d_\sigma
               \right \} .
\nonumber
\end{eqnarray}
Here $\hat{n}_{\sigma}^{d} = d^{\dagger}_{\sigma} d_{\sigma}$, and 
$ c^\dagger_{\e\sigma\alpha}$ creates a conduction electron in 
the lead $\alpha$  of energy $\e$ and density of states
$\rho(\e)$. 

This Hamiltonian is commonly used to model  ultra-small quantum
dots\cite{MeirWingreen1994,NatureGoldhaberGordon1998}. In the absence of the local Coulomb
repulsion $\H_U= U( \sum_{\sigma} \hat{n}^d_\sigma -1)^2/2$, the single
particle problem is
diagonalized exactly in the continuum limit\cite{JongHan2006,Hershfield1993,HanHeary2007,Oguri2007,EnssSchoenhammer2005,LebanonSchillerAndersCB2003}
by  the scattering states operators
\begin{eqnarray}
  \label{eq:scattering-states-operators}
  \gamma^\dagger_{\e \sigma \alpha} &=& c^\dagger_{\e \sigma \alpha} +
  V_\alpha \sqrt{\rho(\e)} G_{0\sigma}^r(\e +i\delta)
 \nonumber \\ && \times 
 \left[
 d^\dagger_{\sigma}  +\sum_{\alpha'} 
 \int d\e' 
 \frac{V_{\alpha'} \sqrt{\rho(\e')}}{\e+i\delta -\e'}
 c^\dagger_{\e'\sigma\alpha'} 
 \right]
\end{eqnarray} 
where $\bar V=\sqrt{V_L^2+V_R^2}$, and the Green function
$G_{0\sigma}^r(z) = \left[z- ( E_d + U/2 -   \sigma H/2) - \bar  V^2
  \int d\e \rho(\e)/(z-\e) \right]^{-1}$.
In the limit of infinitely large leads, the single-particle spectrum
remains unaltered, and these scattering states diagonalize the
Hamiltonian\cite{Hershfield1993} (\ref{eqn:siam})  for $U=0$:
\begin{eqnarray}
  \H^i_0 = \H(U=0) & =& \sum_{\alpha=L,R;\sigma} \int d\e \, \e 
  \gamma^\dagger_{\e\sigma\alpha}\gamma_{\e\sigma\alpha}
  \punkt
\end{eqnarray}
 Hershfield  has shown that the density operator for such a
non-interacting current-carrying quantum system 
retains its Boltzmannian form\cite{Hershfield1993,DoyonAndrei2005}
even at finite bias: 
\begin{eqnarray}
  \hat \rho_0 &= &\frac{e^{-\beta(\H^i_0 -\hat Y_0)}}{\Tr{e^{-\beta(\H^i_0 -\hat Y_0)}}}
 \label{eqn:rho_0}
\, , \,
 \hat Y_0 = \sum_{\alpha\sigma} \mu_\alpha \int d\e \,
 \gamma^\dagger_{\e\sigma\alpha}\gamma_{\e\sigma\alpha} 
\end{eqnarray}
The $\hat Y_0$ operator accounts for the
occupation of the left and right-moving scattering states, and
$\mu_\alpha$ for the different chemical potentials of the leads.

\begin{figure}
  \centering
  \includegraphics[width=70mm]{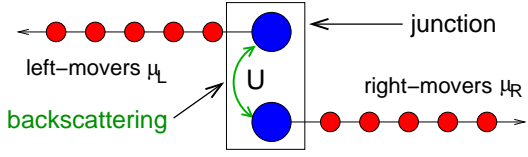}
  \caption{The local $d$-orbital is expanded in
    left-moving and right-moving scattering states. Each
    contributions defines one fictitious local orbital $d_{\sigma\alpha}$ of the junction of the
    scattering-states NRG. The Coulomb repulsion introduces
    backscattering between left and right-movers.}
  \label{fig:1}
\end{figure}

\paragraph{Steady-state NRG:} In order to apply the NRG to such an open
quantum systems, the scattering states $\gamma_{\e\alpha\sigma}$ are
discretized on a logarithmic energy mesh using the NRG
discretization parameter $\Lambda$\cite{BullaCostiPruschke2007}.
In contrary to a closed system, however, each of these single-particle states
carries a finite current. Even for asymmetric coupling, the spectra of
the right and left-movers remains symmetric, and the  total current
vanishes always at zero bias. 

Defining the creation operator for a fictitious left or right-moving
$d_{\sigma\alpha}$-orbital $d^\dagger_{\sigma\alpha} = \bar V \int d\e
\sqrt{\rho(\e)}[  G^r_{0\sigma}(\e+i\delta)]^*
\gamma^\dagger_{\e\sigma\alpha }$, the physical $d$-level can be
decomposed into $d^\dagger_\sigma = r_R d^\dagger_{\sigma R} + r_L
d^\dagger_{\sigma  L}$ by inverting Eq.~(\ref{eq:scattering-states-operators})
and using $r_\alpha = V_\alpha/\bar V$. For $U=0$, the Hamiltonian is diagonal 
in the left and right-movers. We use these
$d_{\sigma\alpha}$-orbitals as starting vector of the Householder
transformation\cite{BullaCostiPruschke2007} mapping the discretized scattering
states continuum onto two semi-infinite Wilson
chains\cite{BullaCostiPruschke2007}, as depicted in
Fig.~\ref{fig:1}. These chains are  almost identical to standard
Wilson chain of  a non-interacting resonant level
model\cite{BullaCostiPruschke2007}. Each fictitious
$d_{\sigma\alpha}$-orbital  consists of a normalized linear
combination of scattering states $\gamma_{\e \sigma \alpha}$: 
no auxiliary degrees of freedom has been  introduced into the problem!

We divide $G^r_{0\sigma}(\e+i\delta)$ into magnitude and phase,
$G^r_{0\sigma}(\e+i\delta)= e^{i\Phi_\sigma(\e)} |G^r_{0\sigma}(\e+i\delta)|$,
and absorb  the energy dependent phase $\Phi_\sigma(\e)$ into the
scattering-states operators $\gamma_{\e\sigma\alpha}$ by a gauge
transformation. Then, the Wilson chains consist only of  purely real
tight-binding parameters. Diagonalizing the proposed
scattering-states Wilson chains yields a  faithful representation of
the steady-state density operator $\hat \rho_0$ for arbitrary bias.

The current operator expanded in scattering states
$\gamma_{\e\sigma\alpha}$  acquires an additional energy dependence
via the scattering-phase shift $\Phi_\sigma(\e)$. In our model
(\ref{eqn:siam}), however, the current remains connected to  the
spectral function $A_d(\w)$ of the retarded non-equilibrium Green
function\cite{Costi97} 
\begin{eqnarray}
  \label{eq:ss-current}
  I(V) &=& \frac{G_0}{e} \sum_\sigma \int_{-\infty}^\infty \, d\w \, 
\left[f(w-\mu_L)-f(w-\mu_R)\right] 
\nonumber 
\\
&& \times
\pi A_{d\sigma}(\w) \Gamma
\end{eqnarray}
in such a  scattering-states formulation  even  for finite
$U$\cite{MeirWingreen1992,Hershfield1993,Oguri2007}.  
$f(\w)$ denotes the Fermi function, 
$G_0 =
(e^2/h) 4\Gamma_L \Gamma_R/(\Gamma_L +\Gamma_R)^2$, $\Gamma_\alpha =
r^2_\alpha \pi \bar V^2 \rho(0)$, 
$\Gamma= \Gamma_L +\Gamma_R$ 
and $\pi
A_{d\sigma}(\w)=-\Im m G^r_{d\sigma}(\w+i\delta)$.


\paragraph{Coulomb interaction:} 
Expanding the operator $\hat n^d_\sigma$ in the orbitals
$d_{\sigma\alpha}$ yields  two contributions: a
density and a backscattering term: 
$\hat n^d_\sigma =  \hat{n}^0_\sigma +\hat O_{\sigma}^{back}$, with 
$\hat{n}^0_\sigma = \sum_\alpha r_\alpha^2
d^\dagger_{\sigma\alpha}d_{\sigma\alpha}$.
 The backscattering term  reads
\begin{eqnarray}
  \hat O_{\sigma}^{back} &=& r_L r_R\left( d^\dagger_{\sigma R}
    d_{\sigma L}
+d^\dagger_{\sigma L} d_{\sigma R}
\right)
\end{eqnarray}
and describes transitions between left and right-movers. 
This term vanishes in the tunnelling regime, where either  $r_L$ or
$r_R$ vanishes.

We will include the full Coulomb interaction into our theory in two
steps. Since $H_U^0$, defined as $  H_U^0 = \frac{U}{2}
\left( \sum_{\sigma} \hat{n}^0_\sigma -1\right)^2$,
commutes with $\hat Y_0$, the steady state density operator $\hat \rho_0$
evolved into $\tilde \rho_0=\exp[-\beta(\H^i-\hat Y_0)]/Z$ with
$\H^i = \H^i_0 +  H_U^0$ proven by the arguments given in
Ref.~\cite{DoyonAndrei2005}.  $\hat O_{\sigma}^{back}$ can be
neglected  in the tunneling regime where $\hat\rho \rightarrow
\tilde\rho_0$. Then, the steady-state spectra  is  completely
determined by a single effective orbital, and the equilibrium spectral
function is recovered.

$\H^i$ marks the new starting point of our theory. The full
Hamiltonian $\H$ of the interacting model differs from $\H^i$ by the
additional backscattering terms. $\H$ does not commute with $\hat
Y_0$, and  the analytical  form of steady-state density operator of
the fully interacting problem is not explicitly
known\cite{Hershfield1993,DoyonAndrei2005}. We obtain a solution\cite{AndersSchiller2005,AndersSchiller2006,AndersNeqGf2008} by
evolving $\tilde \rho_0$ with respect to the full Hamiltonian $\H$
into its steady-state value $\hat \rho_{\infty}  = \lim_{t\to \infty}
e^{-i\H t} \tilde\rho_0 e^{i\H  t}$. In the  current-voltage relation
(\ref{eq:ss-current}), the spectral function $A_{d\sigma}(\w)$ for
$U=0$ is replaced by the  non-equilibrium spectral
function\cite{MeirWingreen1992} calculated with respect to $\hat
\rho_{\infty}$. The details of this algorithm embedding the
calculation of equilibrium spectral
functions\cite{PetersPruschkeAnders2006,WeichselbaumDelft2007}  are
published in Ref.~\cite{AndersNeqGf2008}. 

\paragraph{Results:} 

\begin{figure}
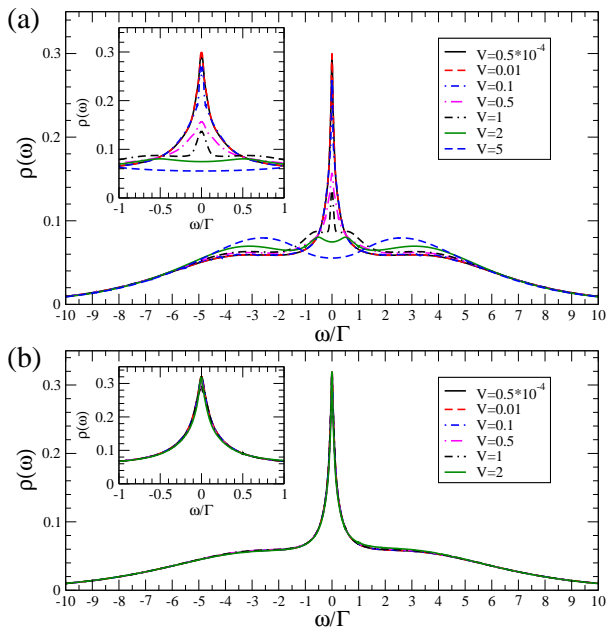

  \centering
  \includegraphics[width=80mm]{fig2a}
  \\
  
  \includegraphics[width=80mm]{fig2b}

  \caption{(color online) Non-equilibrium spectral function for (a) a
    symmetric junction 
    $R=1$ at various values of     finite bias
    voltage $V$, and (b) for a strongly asymmetric junction
    $R=1000$. The insets show  the evolution of the Kondo-resonance. 
    Parameters: $U=8$,    $\e_f=-4$ and $T\to 0$.} 
\label{fig:2}
\end{figure}

All energies are measured in units of $\Gamma = \pi \bar V^2 \rho(0)$;
a constant band width\cite{BullaCostiPruschke2007} of
$\rho(\w)=1/(2D)\Theta(D-|\omega|)$ was used with $D/\Gamma=10$. The
number of retained NRG states was $N_s=2200$; a $\Lambda=4$ was
chosen. The model lacks  channel conservation: only the total charge
and $z$-component of the spin served as quantum 
numbers. We defined $R=\Gamma_L/\Gamma_R$ and always kept
$\Gamma=\Gamma_L+\Gamma_R$ constant. The two chemical
potentials  $\mu_\alpha$ were set to $\mu_L=-r^2_R V$ and $\mu_R=
r^2_L V$ as function of the external source-drain voltage $V$
consistent with a serial resistor model.

The non-equilibrium spectral function for a symmetric
junction is plotted for $U=8$ and different bias $V$ in
Fig.~\ref{fig:2}(a).  Multiple backscattering events cause  gain 
(or lost) of single-particle excitation energy proportional to the
applied bias. The Kondo resonance is destroyed with increasing bias
due to redistribution of spectral weight towards higher energys.
An onset of two weak peaks in the vicinity of the two chemical
potentials remains for $|V|>\Gamma$\cite{HanHeary2007}.  For large
$R\gg 1$ such  backscattering  processes are suppressed. The  spectral
function remains bias-independent. The Kondo 
resonance remains pinned to $\mu_L\to 0$ as depicted in
Fig.\ \ref{fig:2}(b), and  we  recover the tunneling regime.

\begin{figure}
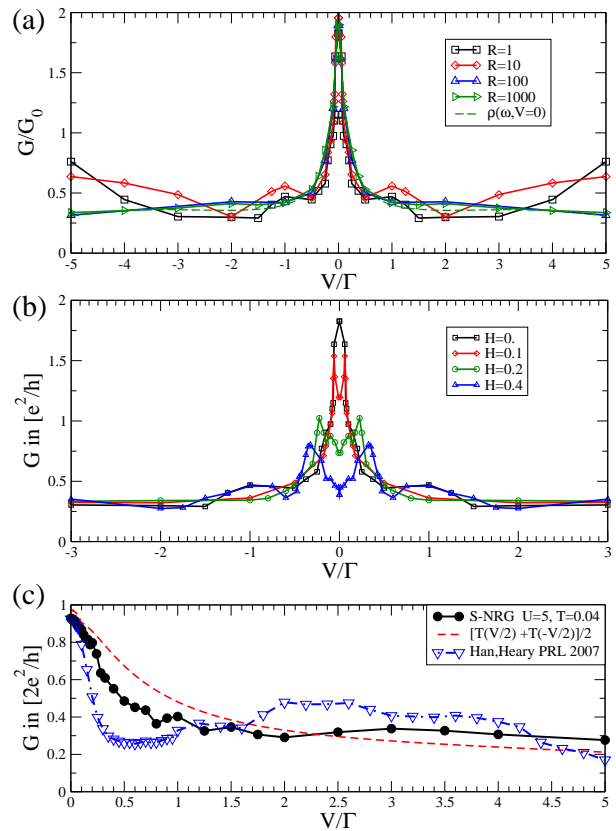

  \centering
   \includegraphics[width=80mm]{fig3a}

   \includegraphics[width=80mm]{fig3b}

   \includegraphics[width=80mm]{fig3c}
   
  \caption{(color online) The differential conductance $G= dI/dV$ as function
    of the bias voltage (a) for different asymmetry factors $R$,
    (b) for different magnetic field
    $H=0,0.1,0.2,0.4$ and $R=1$.
    Parameters: as in Fig.~\ref{fig:2}.
    (c) Comparison between the results for $U=5$ from
    Ref.~\cite{HanHeary2007} and the NRG calculation at
    $T/\Gamma=0.04$ and $R=1$ using z-averaging over 4
    z-values\cite{AndersSchiller2005,AndersSchiller2006}.
  } 
\label{fig:3}
\end{figure}

The differential conductance is plotted for
different asymmetry ratios $R$ in Fig.~\ref{fig:3}(a) using the same
parameters as in Fig.~\ref{fig:2}. With increasing $R$, the
non-equilibrium spectral function is less broadened and, therefore,
$G(V)$ decreases for large bias voltage. Asymptotically, $G$ 
approaches the equilibrium $t$-matrix which is the exact result for
$R\to\infty$ and $T\to 0$.

The effect of an external magnetic field onto the differential
conductance is shown in Fig.~\ref{fig:3}(b).  An increasing magnetic
field splits the zero-bias anomaly which is further suppressed by the
finite  bias in a symmetric junction. This field dependence has been
used in experiments\cite{NatureGoldhaberGordon1998} as 
hallmark for the Kondo physics at low temperatures.

In Fig.~\ref{fig:3}(c), the NRG conductance for $U=5$ is compared to
the result of Ref.~\cite{HanHeary2007}. Both curves agree for low
bias. The NRG result shows a weaker decay of the zero-bias
anomaly with increasing bias with a less pronounce maximum at large bias.
The symmetrized equilibrium t-matrix\cite{BullaCostiPruschke2007} is
added for comparison as dashed line.

\begin{figure}
  \centering
  \includegraphics[width=80mm]{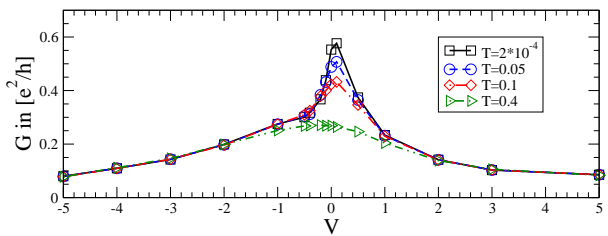}

  \caption{(color online) The differential conductance $G$ as function
    of the bias voltage for different temperatures. Parameters $R=10$,
    $\e_f=-1.5$ and $U=12$.
    }
\label{fig:4}
\end{figure}

The more generic case of an  asymmetric junction with respect 
with a relatively large local Coulomb repulsion is plotted in
Fig.~\ref{fig:4}. The  differential conduction reflects  the
lack of symmetry under source-drain voltage reversal. 
As depicted, the zero-bias peak vanishes
with increasing temperature.

\paragraph{Conclusion:}
A powerful new approach to  the steady-state currents through
nano-devices has been presented. 
We have introduced a  NRG method based on scattering states
to incorporate the correct {\em steady state} boundary condition of
{\em  current carrying  systems}. 
The steady-state density operator\cite{Hershfield1993} of
a non-interacting junction is evolved into the one of the interacting
nano-device using the TD-NRG\cite{AndersSchiller2005}. 
We have established an accurate solution for the strong-coupling regime
and calculated steady-state currents for arbitrary ratios $R$ 
at finite bias. The tunneling regime is included as an exact
limit.
Our approach does not suffer from any current reflection inherent to
numerical simulations of closed quantum systems\cite{Schmitteckert2004}.
We have concentrated on the low-temperature properties of the nano-device,
since the combination of arbitrary bias, large Coulomb repulsion and
finite magnetic field remains the most difficult regime for all
perturbative methods. However, the NRG is equally suitable to
calculate the crossover from the low to the high-temperature
regime as demonstrated in Fig.~\ref{fig:4}. An experimental
hallmark\cite{NatureGoldhaberGordon1998} for Kondo physics, the
splitting of the zero-bias Kondo peak with increasing magnetic field,
is correctly described by our approach for arbitrary temperature, bias
and field strength. 

This theory can be extended to  more complicated multi-orbital
models. Eq.~(\ref{eq:ss-current}) must be modified and requires
more  complex correlation functions.  Since single-particle scattering
states can always be obtained exactly, the construction of the  Wilson
chain parameters is straight forward using the corresponding expansion
of the local degrees of freedom and combining it  with the
transformation used for non-constant density of
states\cite{BullaCostiPruschke2007}.

\begin{acknowledgments}
I acknowledge stimulating discussions with N.~Andrei, J.~Bauer,
G.~Czycholl, Th. Costi, M. Jarrell, H.~Monien, A.~Millis, J.~Kroha,
J. Han for providing the data of Ref.~\cite{HanHeary2007}, Th.~Pruschke, A.~Schiller, P.~Schmitteckert, K.~Schoenhammer, A. Weichselbaum, G.~Uhrig
and the KITP for its  hospitality, at which some of the work has been
carried out.  I also thank  T. Novotny for pointing out
Ref.~\cite{Oguri2007}. This research was supported in parts by the DFG
project AN 275/6-1 and by the National Science Foundation under Grant
No. PHY05-51164. We acknowledge supercomputer support by the
NIC, Forschungszentrum J\"ulich P.No.\ HHB000. 
\end{acknowledgments}


\begin{thebibliography}{30}
\expandafter\ifx\csname natexlab\endcsname\relax\def\natexlab#1{#1}\fi
\expandafter\ifx\csname bibnamefont\endcsname\relax
  \def\bibnamefont#1{#1}\fi
\expandafter\ifx\csname bibfnamefont\endcsname\relax
  \def\bibfnamefont#1{#1}\fi
\expandafter\ifx\csname citenamefont\endcsname\relax
  \def\citenamefont#1{#1}\fi
\expandafter\ifx\csname url\endcsname\relax
  \def\url#1{\texttt{#1}}\fi
\expandafter\ifx\csname urlprefix\endcsname\relax\def\urlprefix{URL }\fi
\providecommand{\bibinfo}[2]{#2}
\providecommand{\eprint}[2][]{\url{#2}}

\bibitem[{\citenamefont{Kastner}(1992)}]{KastnerSET1992}
\bibinfo{author}{\bibfnamefont{M.~A.} \bibnamefont{Kastner}},
  \bibinfo{journal}{Rev. Mod. Phys.} \textbf{\bibinfo{volume}{64}},
  \bibinfo{pages}{849} (\bibinfo{year}{1992}).

\bibitem[{\citenamefont{Goldhaber-Gordon
  et~al.}(1998)\citenamefont{Goldhaber-Gordon, Shtrikman, Mahalu,
  Abusch-Magder, Meirav, and Kastner}}]{NatureGoldhaberGordon1998}
\bibinfo{author}{\bibfnamefont{D.}~\bibnamefont{Goldhaber-Gordon}},
   et al.~\bibinfo{journal}{Nature} \textbf{\bibinfo{volume}{391}},
  \bibinfo{pages}{156} (\bibinfo{year}{1998}).

\bibitem[{\citenamefont{van~der Wiel et~al.}(2000)\citenamefont{van~der Wiel,
  Franceschi, Elzerman, Tarucha, and Kouvenhoven}}]{Kouwenhoven2000}
\bibinfo{author}{\bibfnamefont{W.~G.} \bibnamefont{van~der Wiel}},
 et al.~\bibinfo{journal}{Science} \textbf{\bibinfo{volume}{289}},
  \bibinfo{pages}{2105} (\bibinfo{year}{2000}).

\bibitem[{\citenamefont{Bulla et~al.}(2008)\citenamefont{Bulla, Costi, and
  Pruschke}}]{BullaCostiPruschke2007}
\bibinfo{author}{\bibfnamefont{R.}~\bibnamefont{Bulla}},
  \bibinfo{author}{\bibfnamefont{T.~A.} \bibnamefont{Costi}}, \bibnamefont{and}
  \bibinfo{author}{\bibfnamefont{T.}~\bibnamefont{Pruschke}},
  \bibinfo{journal}{Rev.~Mod.~Phys.} \textbf{\bibinfo{volume}{80}},
  \bibinfo{pages}{395} (\bibinfo{year}{2008}).

\bibitem[{\citenamefont{Wingreen and Meir}(1994)}]{MeirWingreen1994}
\bibinfo{author}{\bibfnamefont{N.~S.} \bibnamefont{Wingreen}} \bibnamefont{and}
  \bibinfo{author}{\bibfnamefont{Y.}~\bibnamefont{Meir}},
  \bibinfo{journal}{Phys.~Rev.~B} \textbf{\bibinfo{volume}{49}},
  \bibinfo{pages}{11040} (\bibinfo{year}{1994}).

\bibitem[{\citenamefont{K\"onig and Schoeller}(1998)}]{KoenigSchoeller98}
\bibinfo{author}{\bibfnamefont{J.}~\bibnamefont{K\"onig}} \bibnamefont{and}
  \bibinfo{author}{\bibfnamefont{H.}~\bibnamefont{Schoeller}},
  \bibinfo{journal}{Phys.~Rev.~Lett.} \textbf{\bibinfo{volume}{81}},
  \bibinfo{pages}{3511} (\bibinfo{year}{1998}).

\bibitem[{\citenamefont{Rosch et~al.}(2003)\citenamefont{Rosch, Paaske, Kroha,
  and W\"olfle}}]{RoschPaaskeKrohaWoelfe2003}
\bibinfo{author}{\bibfnamefont{A.}~\bibnamefont{Rosch}},
  \bibinfo{author}{\bibfnamefont{J.}~\bibnamefont{Paaske}},
  \bibinfo{author}{\bibfnamefont{J.}~\bibnamefont{Kroha}}, \bibnamefont{and}
  \bibinfo{author}{\bibfnamefont{P.}~\bibnamefont{W\"olfle}},
  \bibinfo{journal}{Phys.~Rev.~Lett.} \textbf{\bibinfo{volume}{90}},
  \bibinfo{pages}{076804} (\bibinfo{year}{2003}).

\bibitem[{\citenamefont{Gezzi et~al.}(2007)\citenamefont{Gezzi, Pruschke, and
  Meden}}]{GezziPruschkeMeden2007}
\bibinfo{author}{\bibfnamefont{R.}~\bibnamefont{Gezzi}},
  \bibinfo{author}{\bibfnamefont{T.}~\bibnamefont{Pruschke}}, \bibnamefont{and}
  \bibinfo{author}{\bibfnamefont{V.}~\bibnamefont{Meden}},
  \bibinfo{journal}{Phys. Rev. B} \textbf{\bibinfo{volume}{75}},
  \bibinfo{pages}{045324} (\bibinfo{year}{2007}).

\bibitem[{\citenamefont{Keldysh}(1965)}]{Keldysh65}
\bibinfo{author}{\bibfnamefont{L.~V.} \bibnamefont{Keldysh}},
  \bibinfo{journal}{Sov. Phys. JETP} \textbf{\bibinfo{volume}{20}},
  \bibinfo{pages}{1018} (\bibinfo{year}{1965}).

\bibitem[{\citenamefont{Schiller and Hershfield}(1995)}]{SchillerHershfield95}
\bibinfo{author}{\bibfnamefont{A.}~\bibnamefont{Schiller}} \bibnamefont{and}
  \bibinfo{author}{\bibfnamefont{S.}~\bibnamefont{Hershfield}},
  \bibinfo{journal}{Phys.~Rev.~B} \textbf{\bibinfo{volume}{51}},
  \bibinfo{pages}{12896} (\bibinfo{year}{1995}).

\bibitem[{\citenamefont{Kehrein}(2005)}]{Kehrein2005}
\bibinfo{author}{\bibfnamefont{S.}~\bibnamefont{Kehrein}},
  \bibinfo{journal}{Phys.~Rev.~Lett.} \textbf{\bibinfo{volume}{95}},
  \bibinfo{pages}{056602} (\bibinfo{year}{2005}).

\bibitem[{\citenamefont{Cuniberti et~al.}(2005)\citenamefont{Cuniberti, Fagas,
  and Richter}}]{MolecularElectronicsBook2005}
\bibinfo{editor}{\bibfnamefont{G.}~\bibnamefont{Cuniberti}},
  \bibinfo{editor}{\bibfnamefont{G.}~\bibnamefont{Fagas}}, \bibnamefont{and}
  \bibinfo{editor}{\bibfnamefont{K.}~\bibnamefont{Richter}}, eds.,
  \emph{\bibinfo{title}{Introducing Molecular Electronics}}, vol.
  \bibinfo{volume}{680} of \emph{\bibinfo{series}{Lecture Notes in Physics}}
  (\bibinfo{publisher}{Springer}, \bibinfo{address}{Berlin and Heidelberg},
  \bibinfo{year}{2005}).

\bibitem[{\citenamefont{Welack et~al.}(2006)\citenamefont{Welack, Schreiber,
  and Kleinekathoefer}}]{welack-2006-124}
\bibinfo{author}{\bibfnamefont{S.}~\bibnamefont{Welack}},
  \bibinfo{author}{\bibfnamefont{M.}~\bibnamefont{Schreiber}},
  \bibnamefont{and}
  \bibinfo{author}{\bibfnamefont{U.}~\bibnamefont{Kleinekathoefer}},
  \bibinfo{journal}{J. Chem. Phys.} \textbf{\bibinfo{volume}{124}},
  \bibinfo{pages}{044712} (\bibinfo{year}{2006}).

\bibitem[{\citenamefont{Han}(2006)}]{JongHan2006}
\bibinfo{author}{\bibfnamefont{J.~E.} \bibnamefont{Han}},
  \bibinfo{journal}{Phys.~Rev.~B} \textbf{\bibinfo{volume}{73}},
  \bibinfo{pages}{125319} (\bibinfo{year}{2006}).

\bibitem[{\citenamefont{Hershfield}(1993)}]{Hershfield1993}
\bibinfo{author}{\bibfnamefont{S.}~\bibnamefont{Hershfield}},
  \bibinfo{journal}{Phys.~Rev.~Lett.} \textbf{\bibinfo{volume}{70}},
  \bibinfo{pages}{2134} (\bibinfo{year}{1993}).

\bibitem[{\citenamefont{Han and Heary}(2007)}]{HanHeary2007}
\bibinfo{author}{\bibfnamefont{J.~E.} \bibnamefont{Han}} \bibnamefont{and}
  \bibinfo{author}{\bibfnamefont{R.~J.} \bibnamefont{Heary}},
  \bibinfo{journal}{Phys.~Rev.~Lett.} \textbf{\bibinfo{volume}{99}},
  \bibinfo{pages}{236808} (\bibinfo{year}{2007}).

\bibitem[{\citenamefont{Oguri}(2007)}]{Oguri2007}
\bibinfo{author}{\bibfnamefont{A.}~\bibnamefont{Oguri}},
  \bibinfo{journal}{Phys.~Rev.~B} \textbf{\bibinfo{volume}{75}},
  \bibinfo{pages}{035302} (\bibinfo{year}{2007}).

\bibitem[{\citenamefont{Doyon and Andrei}(2006)}]{DoyonAndrei2005}
\bibinfo{author}{\bibfnamefont{B.}~\bibnamefont{Doyon}} \bibnamefont{and}
  \bibinfo{author}{\bibfnamefont{N.}~\bibnamefont{Andrei}},
  \bibinfo{journal}{Phys.~Rev.~B} \textbf{\bibinfo{volume}{73}},
  \bibinfo{pages}{245326} (\bibinfo{year}{2006}).

\bibitem[{\citenamefont{Mehta and Andrei}(2006)}]{MethaAndrei2005}
\bibinfo{author}{\bibfnamefont{P.}~\bibnamefont{Mehta}} \bibnamefont{and}
  \bibinfo{author}{\bibfnamefont{N.}~\bibnamefont{Andrei}},
  \bibinfo{journal}{Phys.~Rev.~Lett.} \textbf{\bibinfo{volume}{96}},
  \bibinfo{pages}{216802} (\bibinfo{year}{2006}).

\bibitem[{\citenamefont{Anders and Schiller}(2005)}]{AndersSchiller2005}
\bibinfo{author}{\bibfnamefont{F.~B.} \bibnamefont{Anders}} \bibnamefont{and}
  \bibinfo{author}{\bibfnamefont{A.}~\bibnamefont{Schiller}},
  \bibinfo{journal}{Phys.~Rev.~Lett.} \textbf{\bibinfo{volume}{95}},
  \bibinfo{pages}{196801} (\bibinfo{year}{2005}).

\bibitem[{\citenamefont{Anders and Schiller}(2006)}]{AndersSchiller2006}
\bibinfo{author}{\bibfnamefont{F.~B.} \bibnamefont{Anders}} \bibnamefont{and}
  \bibinfo{author}{\bibfnamefont{A.}~\bibnamefont{Schiller}},
  \bibinfo{journal}{Phys.~Rev.~B} \textbf{\bibinfo{volume}{74}},
  \bibinfo{pages}{245113} (\bibinfo{year}{2006}).

\bibitem[{\citenamefont{Anders}(2008)}]{AndersNeqGf2008}
\bibinfo{author}{\bibfnamefont{F.~B.} \bibnamefont{Anders}},
  \bibinfo{journal}{J. Phys.: Condens. Matter} \textbf{\bibinfo{volume}{20}},
  \bibinfo{pages}{195216} (\bibinfo{year}{2008}).

\bibitem[{\citenamefont{Schollw\"ock}(2005)}]{SchollwoeckDMRG2005}
\bibinfo{author}{\bibfnamefont{U.}~\bibnamefont{Schollw\"ock}},
  \bibinfo{journal}{Rev. Mod. Phys.} \textbf{\bibinfo{volume}{77}},
  \bibinfo{pages}{259} (\bibinfo{year}{2005}).

\bibitem[{\citenamefont{Schmitteckert}(2004)}]{Schmitteckert2004}
\bibinfo{author}{\bibfnamefont{P.}~\bibnamefont{Schmitteckert}},
  \bibinfo{journal}{Phys.~Rev.~B} \textbf{\bibinfo{volume}{70}},
  \bibinfo{pages}{121302(R)} (\bibinfo{year}{2004}).

\bibitem[{\citenamefont{Lebanon et~al.}(2003)\citenamefont{Lebanon, Schiller,
  and Anders}}]{LebanonSchillerAndersCB2003}
\bibinfo{author}{\bibfnamefont{E.}~\bibnamefont{Lebanon}},
  \bibinfo{author}{\bibfnamefont{A.}~\bibnamefont{Schiller}}, \bibnamefont{and}
  \bibinfo{author}{\bibfnamefont{F.~B.} \bibnamefont{Anders}},
  \bibinfo{journal}{Phys.~Rev.~B} \textbf{\bibinfo{volume}{68}},
  \bibinfo{pages}{155301} (\bibinfo{year}{2003}).

\bibitem[{\citenamefont{Enss et~al.}(2005)\citenamefont{Enss, Meden,
  Andergassen, Barnabe-Theriault, Metzner, and
  Schoenhammer}}]{EnssSchoenhammer2005}
\bibinfo{author}{\bibfnamefont{T.}~\bibnamefont{Enss}},
 et al.~ \bibinfo{journal}{Phys. Rev. B} \textbf{\bibinfo{volume}{71}},
  \bibinfo{pages}{155401} (\bibinfo{year}{2005}).

\bibitem[{\citenamefont{Costi}(1997)}]{Costi97}
\bibinfo{author}{\bibfnamefont{T.~A.} \bibnamefont{Costi}},
  \bibinfo{journal}{Phys.~Rev.~B} \textbf{\bibinfo{volume}{55}},
  \bibinfo{pages}{3003} (\bibinfo{year}{1997}).

\bibitem[{\citenamefont{Meir and Wingreen}(1992)}]{MeirWingreen1992}
\bibinfo{author}{\bibfnamefont{Y.}~\bibnamefont{Meir}} \bibnamefont{and}
  \bibinfo{author}{\bibfnamefont{N.~S.} \bibnamefont{Wingreen}},
  \bibinfo{journal}{Phys.~Rev.~Lett.} \textbf{\bibinfo{volume}{68}},
  \bibinfo{pages}{2512} (\bibinfo{year}{1992}).

\bibitem[{\citenamefont{Peters et~al.}(2006)\citenamefont{Peters, Pruschke, and
  Anders}}]{PetersPruschkeAnders2006}
\bibinfo{author}{\bibfnamefont{R.}~\bibnamefont{Peters}},
  \bibinfo{author}{\bibfnamefont{T.}~\bibnamefont{Pruschke}}, \bibnamefont{and}
  \bibinfo{author}{\bibfnamefont{F.~B.} \bibnamefont{Anders}},
  \bibinfo{journal}{Phys.~Rev.~B} \textbf{\bibinfo{volume}{74}},
  \bibinfo{pages}{245114} (\bibinfo{year}{2006}).

\bibitem[{\citenamefont{Weichselbaum and von
  Delft}(2007)}]{WeichselbaumDelft2007}
\bibinfo{author}{\bibfnamefont{A.}~\bibnamefont{Weichselbaum}}
  \bibnamefont{and} \bibinfo{author}{\bibfnamefont{J.}~\bibnamefont{von
  Delft}}, \bibinfo{journal}{Phys.~Rev.~Lett.} \textbf{\bibinfo{volume}{99}},
  \bibinfo{pages}{076402} (\bibinfo{year}{2007}).

\end{thebibliography}

\end{document}